\documentclass[twocolumn,showpacs,preprintnumbers,amsmath,amssymb]{revtex4}
\usepackage{graphicx}% Include figure files
\usepackage{dcolumn}% Align table columns on decimal point
\usepackage{bm}% bold math
\usepackage[tight]{subfigure}

\newcommand{\half}{\textstyle \frac{1}{2}}

\begin{document}

\title{
  Spin quantum tunneling in single molecular magnets: 
  fingerprints in transport spectroscopy of current and noise
}
\author{C. Romeike}
\author{M. R. Wegewijs}
\author{H. Schoeller}
\affiliation{Institut f\"ur Theoretische Physik A, RWTH Aachen,\\ 52056 Aachen,  Germany}

\date{\today}

\begin{abstract}
We demonstrate that transport spectroscopy of \emph{single} molecular magnets
shows signatures of quantum tunneling at low 
temperatures. We find current and noise oscillations as function of bias
voltage due to a weak violation of spin selection rules by quantum
tunneling processes. The interplay with Boltzmann suppression factors
leads to fake resonances with temperature-dependent position which do not
correspond to any charge excitation energy. Furthermore, we find that quantum
tunneling can completely suppress transport if the  transverse
anisotropy has a high symmetry.
\end{abstract}

\maketitle

\begin{figure}
 \includegraphics[scale=0.17]{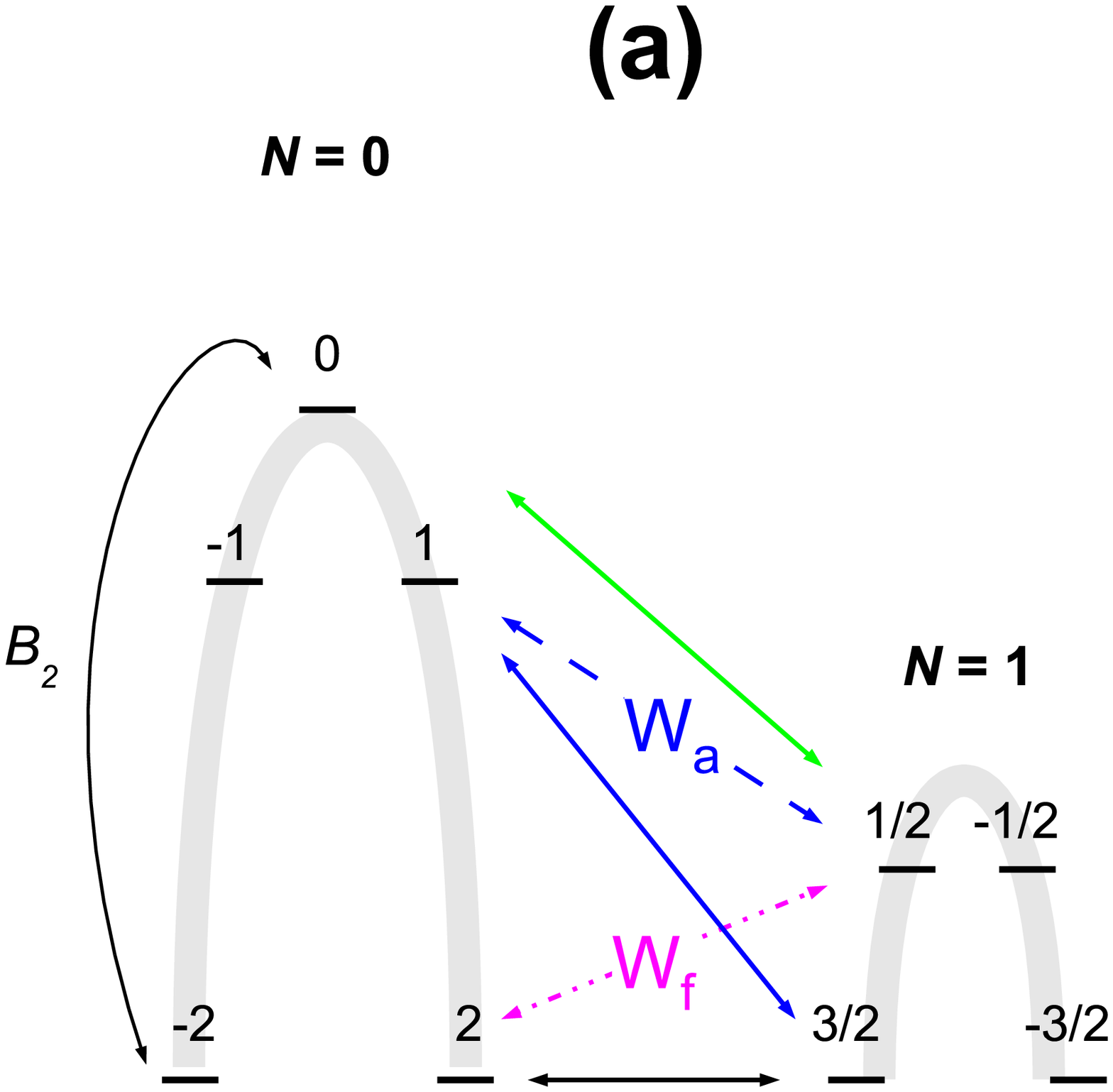}
 \includegraphics[scale=0.17]{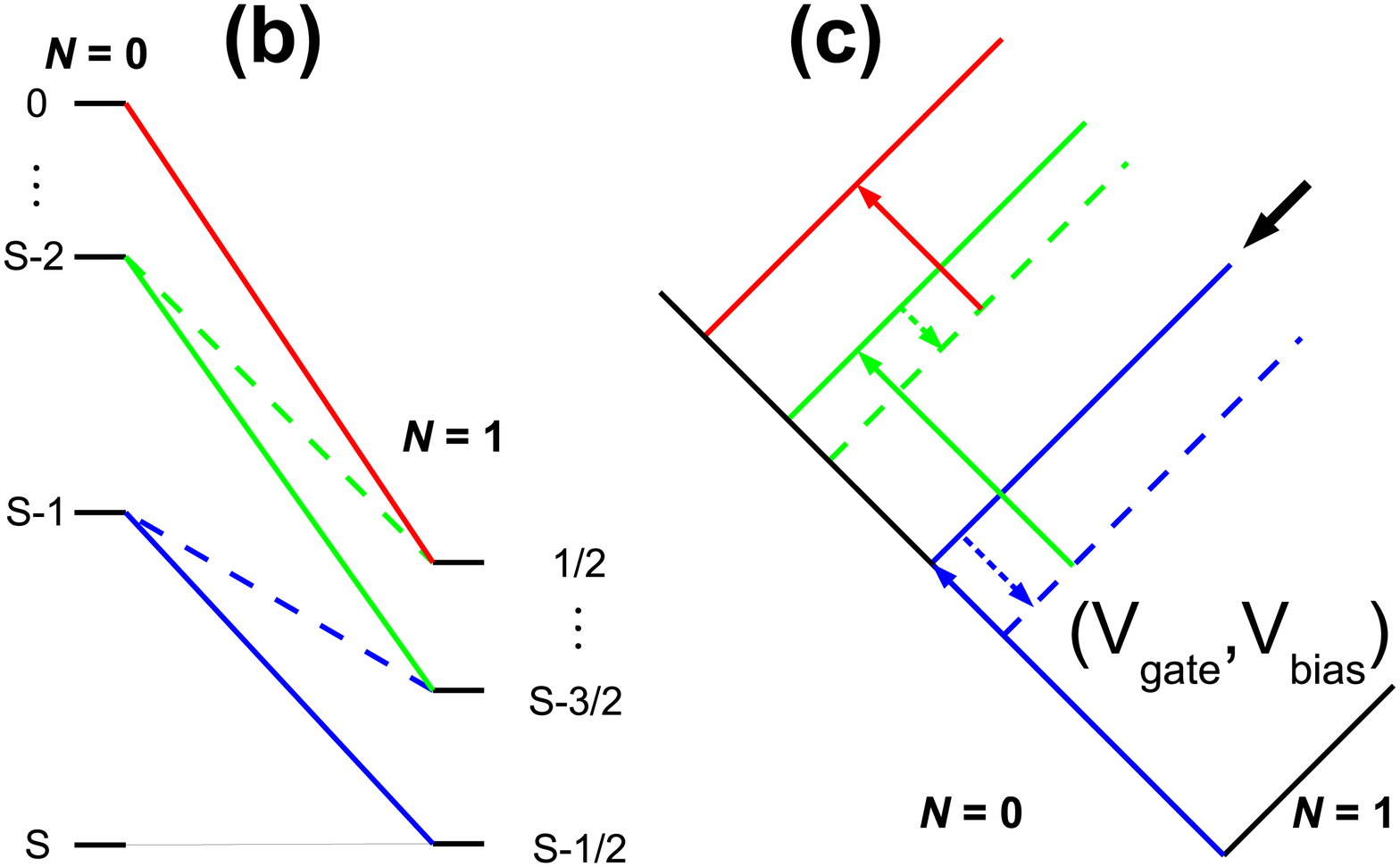}
  \caption{  
    \label{fig:model}
    (a) Magnetic excitation spectra for two charge states with spins 
    $S^{(0)}=2 > S^{(1)}=3/2$ and  $D^{(0)} > D^{(1)}$. 
    Since typically $B_{2n} \ll D^{(N)}$ we label the eigenstates by
    the approximately good quantum number $M$. 
    The dot-dashed line is a spin-forbidden transition, all 
    others are spin-allowed. 
    (b) Energy levels and spin-allowed transitions 
    for $S^{(0)}=S > S^{(1)}=S-{1\over 2}$. 
    (c) Electron-addition excitations in 
    $(V_{\text{gate}},V_{\text{bias}})$ stability diagram.
    Arrows indicate how to construct the diagram going along the 
    zig-zag path in (b). 
    Thick/thin lines indicate visible/hidden transitions. 
    A transition is hidden when the initial state is not yet occupied
    (by other processes) at the transition energy. 
  } 
\end{figure}

\begin{figure}
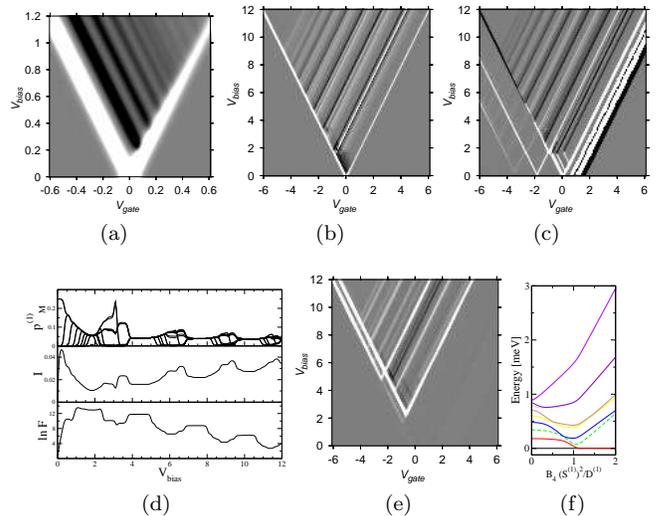

  \subfigure[]{\includegraphics[width=0.32\linewidth]{fig3a.ps}\label{fig:bigS_zoom}}
  \subfigure[]{\includegraphics[width=0.32\linewidth]{fig3b.ps}\label{fig:bigS_didv}}
  \subfigure[]{\includegraphics[width=0.32\linewidth]{fig3c.ps}\label{fig:bigS_noise}}
  \subfigure[]{\includegraphics[width=0.4\linewidth]{fig3d.eps}\label{fig:bigS_curves}}
  \subfigure[]{\includegraphics[width=0.32\linewidth]{fig3e.ps}\label{fig:bigS_block}}
  \subfigure[]{\includegraphics[width=0.19\linewidth]{fig3f.eps}\label{fig:bigS_level}}
    \caption{
    \label{fig:bigS}
    Transport oscillations induced by QTM.
    Parameters:
    $S^{(0)}=10, S^{(1)}=9\half$,
    $D^{(0)}=0.1,D^{(1)}=0.01$, 
    $B_2=2 \times 10^{-3}$ and $T=0.015$.
    (a) $dI/dV_{\text{bias}}$ for small bias: the $N=1$ ``flat'' parabola is mapped
    out by NDC excitations.
    (b) $dI/dV_{\text{bias}}$ for  large bias: the $N=0$ excitations give rise to
    positive and negative differential conductance.
    (c) $d \ln F/dV_{\text{bias}}$: 
    fake resonance lines correspond to noise suppression (black)
    lines and terminate at the Coulomb diamond edge.
    (d) Occupations of the $N=1$ states
    $p^{(1)}_{M}$ together with $ \ln F$ and $I$
    as function of $V_\text{bias}$ for $V_\text{gate}=0$. 
    (e) Current suppression due to high-symmetry QTM:
    $B_4=2 \times 10^{-4}, B_2=0$.
   (f) Spin excitation spectrum for the charged state $N=1$: the
    value $B_4 S^2/D^{(1)}$ used in (e) lies beyond the level crossing at
    approximately $1$. Only the 2nd excited state has a non-negligible
    admixture of the maximal $M'=\pm S^{(1)}$ state which is required
    for transport.
  }
\end{figure}

\emph{Introduction.} Single molecular magnets (SMM) have become 
famous in the last decade for showing the quantum tunneling of a 
single magnetic moment (QTM) on a macroscopic 
scale~\cite{Sessoli93,Friedman96,Hern96,Thomas96,Leuenberger00,Pohjola00}. 
These molecules are characterized by 
a large spin $S>\half$, a large magnetic anisotropy barrier (MAB)
and anisotropy terms which allow this spin to tunnel through the
barrier. 
The anisotropy is due to spin-orbit effects on the metal-ions whose
spins couple to form the large magnetic moment. 
Magnetic hysteresis, associated with QTM, was observed at temperatures below the
MAB~\cite{Sessoli93,Friedman96} for ensembles of molecules in a single crystal.   
Recently, Cornia et al.~\cite{Cornia03} were able to immobilize \emph{single}
SMMs on a gold surface through modification of their ligands 
while preserving the magnetic properties of the core. 
Using this technique Heersche et al.~\cite{Heersche05} were able to
establish a 3-terminal electrical contact and measure the transport
through the well-known SMM Mn12, see also~\cite{Jo06}. \\
In this Letter we show that transport spectroscopy of single molecular magnets 
can reveal specific features being fingerprints of spin quantum tunneling.
Even when the anisotropy terms which cause QTM have a small effect on 
the energy spectrum they lead to significant changes in the  
\emph{non-equilibrium occupations of the magnetic states} 
since they allow for a violation of spin-selection rules for electron-tunneling. 
As a consequence, QTM leads to an oscillatory behavior of 
the current and shot-noise with increasing bias voltage. 
Specifically, the interplay of several small rates (quantum tunneling induced rates
and rates suppressed by Boltzmann factors) leads to negative
differential conductance and, most strikingly, to the occurrence of so-called
fake resonances which do \emph{not} correspond to any charge excitation
energy. The fake resonance's position depends on temperature, and
allows a clear experimental identification of quantum tunneling processes.
Furthermore, we show that high symmetry (due to the molecular structure) 
QTM can give rise to a complete current suppression.\\
\emph{Theory.} We analyze a minimal model that combines the well-known effective spin
Hamiltonian description of SMMs~\cite{Pederson99,Pederson00,Kortus02,Boukhvalov02,Park04add}
 with the standard tunneling Hamiltonian for the coupling to metallic electrodes. Due to the
high charging energy it is sufficient to consider only two charge states
($N=0,1$) with a magnetic excitation spectrum 
$H^{(N)}=H^{(N)}_{\text{MAB}}+H^{(N)}_{ \text{QTM}}$, where
\begin{eqnarray}
    \label{eq:z}
     H^{(N)}_{\text{MAB}} &=&  
     - D^{(N)}
       (\hat S_z)^2 \\
    \label{eq:t}
     H^{(N)}_{ \text{QTM}} &=& - \frac{1}{2} \sum_{n=1,2} B_{2n}^{}
     \left[ (\hat{S}_{+}^2)^{n}+(\hat S_{-}^2)^{n} \right].
\end{eqnarray}
(We employ units $\hbar = e = k_B =1$ and energy units of meV).
For each charge state $N$ the spin has a definite value
$S^{(N)}$ and spin projection $|M| \leq S^{(N)}$ which is maximal in the ground state.
The anisotropy terms arise due to spin-orbit interaction on the molecule
and break rotational invariance in spin-space. 
The lowest order easy-axis anisotropy in Eq.~(\ref{eq:z}) 
defines the preferred axis in space along which we quantize the
spin ($z$-axis). 
The eigenstates  $|N, S, M \rangle $ of Eq.~(\ref{eq:z}) have an inverted parabolic
energy dependence depicted in Fig. \ref{fig:model}(a).
Higher order corrections to the magnetic anisotropy barrier are
not essential here. 
It is known experimentally
\cite{Sessoli1993b,Takeda1997,Aubin1999,Soler00,Soler01,Kuroda-Sowa01,Coronado04} 
and theoretically \cite{Park04add} that 
the anisotropy constants $D^{(N)}$ depend on the charge state. 
The transverse anisotropy, Eq.~(\ref{eq:t}), 
accounts for deviations from purely axial symmetry. 
We consider either a second or fourth ($n=1$ or $2$) order term 
which allows for tunneling of the spin between states with $M$ 
values differing by 2$n$. (It is convenient here to deviate from
the conventional notation $E=B_2$ and $C=B_4$).
Since a charge-dependent QTM induces only small corrections in the 
spectrum of the molecule $B_{2n}$ is taken as charge-independent. 
Transport through SMMs provides information on the magnetic 
structure in more than one charge state of the molecule. 
Therefore we investigate the basic possible combinations of the magnetic
parameter values for charged SMMs which are scarcely known. 
Also, in a single-molecule junction they may 
change due to mechanical and electrostatic effects.
Below we select our values from the typical range
$ D^{(N)} \sim 0.01-0.1$meV, 
 $B^{}_{2} \sim 10^{-3}-10^{-7} $meV and
 $B^{}_{4} \sim 10^{-4}-10^{-7} $meV 
for which magnetic excitations can be resolved in the transport 
at electron temperatures below 1~K.
Since QTM weakly affects the energy spectrum
we will label eigenstates of $H^{(N)}$
by the approximately  good quantum number $M$, 
i.e. state $| N,S,M \rangle$ has the largest contribution. \\
The electrodes $r=L,R$  are described as electron reservoirs with electrochemical 
potential $\mu \pm V_\text{bias}/2$ and a constant density of states $\rho$: 
$H_{\text{res}} =
 \sum_{r k \sigma} (\epsilon_{k \sigma r} - \mu_r) c^{\dag}_{r k
   \sigma} c_{r k \sigma}$.
The tunneling term
$H_{\text{mol-res}} =
\sum_{r k j  \sigma} t_{j} d^{\dag}_{j \sigma} c_{r k \sigma} + h.c. $
describes charge transfer between electrode and molecule (symmetric
tunneling barriers).
Here $d^{\dag}_{j \sigma}$ adds an electron with spin $\sigma$ to a single-particle
orbital on the molecule.
The coupling to a gate electrode is included in a shift of the
molecular energies, such that the charge degeneracy point
is at zero bias ($V_{\text{bias}}$) and gate voltage
($V_{\text{gate}}$). 
For weak tunneling, we use a standard master equation approach
to calculate the non-equilibrium occupations
of the molecular states, the current and the shot noise \cite{thielmann04a}.
The rates in this master equation are calculated in
golden rule approximation. For the transition $s_2\rightarrow s_1$ 
($s_i$ being two eigenstates of $H^{(N_i)}$ with energy $E_i$), we obtain the total
rate $W_{s_1,s_2} = \sum_{r}  W_{s_1,s_2}^{r,+}+W_{s_1,s_2}^{r,-}$ with
the tunneling-in rate
$
  W_{s_1,s_2}^{r,+}  = 2\pi\rho
  \sum_{ \sigma}    f_{r}(E_{s_1}-E_{s_2})
  | T^{\sigma}_{s_1s_2} |^{2}
$
and the tunneling-out rate
$
 W_{s_1,s_2}^{r,-}  = 2\pi\rho
 \sum_{\sigma} (1-f_{r}(E_{s_1}-E_{s_2}))
  | T^{\sigma}_{s_2s_1} |^{2}
$. Here,
$f_r(E)=(e^{(E-\mu_r)/T}+1)^{-1}$ is the Fermi function of reservoir $r$, 
$T$ denotes the temperature, and 
$T^{\sigma}_{s_1s_2} 
=  \sum_{j} t_{j}  \langle  s_1 |  d^{\dag}_{j \sigma} |  s_2 \rangle$.
These tunnel matrix elements incorporate the spin selection rules
and their violation for finite QTM. Without QTM, the eigenstates are
given by $|s_i\rangle=|N_i,S_i,M_i\rangle$, and the tunnel matrix
elements fulfill obviously the spin selection rule $|S_1-S_2|=1/2$ and
$|M_1-M_2|=1/2$, in addition to $|N_1-N_2|=1$. For weak QTM, we 
decompose the states $s_i$ into
a linear combination of $|N_i, S_i, M_i' \rangle$ states, the one
with largest contribution being $M_i' = M_i$. 
Inserting this expansion into $T^{\sigma}_{s_1s_2}$ leads to a 
summation of matrix elements with terms
$\langle N_1,  S_1,  M_1' | \sum_j t_j d^{\dag}_{j\sigma} |N_2, S_2, M_2' \rangle$.
Using the Wigner-Eckart theorem, each of these matrix element can be 
factorized into an $M$-dependent Clebsch-Gordan (CG) coefficient and 
a common constant $c_j$. Each individual CG-coefficient fulfils 
$|S_1-S_2| =1/2$ and $|M_1'-M_2'|=1/2$. Obviously
the overall spin selection rule $|M_1-M_2|=1/2$ can be weakly violated, 
i.e. the corresponding rate is smaller by roughly a factor 
$(B_{2n}/D^{(N)})^2$ compared to the rates fulfilling the 
overall spin selection rule. 
The constants $c_j$ are incorporated into a factor 
$\Gamma = 2\pi\rho|\sum_j t_j c_j|^2$ common to all rates and drop 
out of the problem, except for setting the absolute current and noise
scale. We note that, in contrast to customary spin-blockade
physics~\cite{Weinmann95}, a complete elimination of the spin
projection $M$ from the transport problem is not possible due to the
MAB and QTM.
The master equation approach correctly accounts for both the non-equilibrium 
induced by the electron tunneling at finite bias voltage and the thermal
excitation of molecular spin-states. 
The life-time of the latter is also limited by other relaxation processes
(spin-phonon interaction, nuclear spins, etc.) which are typically~\cite{Wernsdorfer00}
slower than electron tunneling processes (time $\lesssim 1$ ns) and are 
therefore neglected. 
Furthermore, the spin-phonon interaction may be hindered since 
the phonon spectrum for a single molecule is expected to
be less dense than in a bulk system. 
\\
\begin{figure}
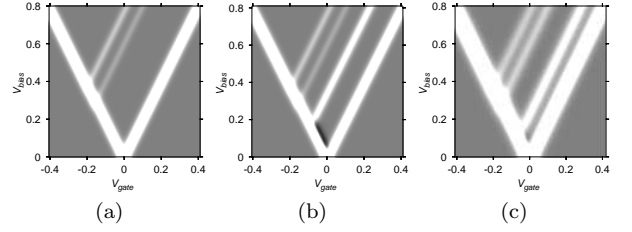

  \subfigure[][]{\includegraphics[scale =0.2]{fig2a_new.ps}}
  \subfigure[][]{\includegraphics[scale =0.2]{fig2c.ps}}
  \subfigure[][]{\includegraphics[scale =0.2]{fig2e.ps}}
  \caption{
    \label{fig:smallS} 
    $dI/dV_{\text{bias}}$ in gray-scale 
    (gray= zero, white/black = positive/negative) 
    as function of $V_\text{gate}$ and $V_\text{bias}$. 
    Parameters:
    $S^{(0)}=2,  S^{(1)}=3/2$, $D^{(0)}=0.1, D^{(1)}=0.01$
     and $T=0.01$. 
    (a) No QTM: $B_2=0$.
    (b) QTM $B_2=2 \times 10^{-5}$
    (c) Same as (b) except for higher $T=0.015$.  
  }
\end{figure}
\emph{Fake resonances and oscillations.} To illustrate the effect of 
the QTM on the transport we first explain the conductance map 
for small spins ($S^{(0)}=2, S^{(1)}=3/2$) and contrast the cases 
$B_2=0$ and $B_2 \neq 0$. 
For $B_2=0$ the differential conductance map is plotted in
Fig.~\ref{fig:smallS}(a) and we discuss the resonance lines running upward.
Starting at the charge degeneracy point ($V_{gate}=V_{bias}=0$) and
increasing the bias voltage the current initially sets on due
to the ground state transitions $M= \pm 2 \leftrightarrow
M'= \pm 3/2$, see Fig. \ref{fig:model}(a).
Increasing the bias voltage further brings the transition 
$M= \pm 2 \leftrightarrow M'= \pm 1/2$ into the transport energy
window, without any effect on the current:
the rate for the process vanishes due to spin selection rules,
$W_f=0$ [dot-dashed in Fig. \ref{fig:model}(a)].
A reservoir spin-1/2 electron can not couple two molecular states with
$|\Delta M|>1/2$. Therefore this resonance is hidden~\cite{Note2,Timm05}. 
The current only increases when the transition $M= \pm 1 \leftrightarrow 
M'= \pm 3/2$ becomes energetically allowed. At this resonance
all states except $N=0,M=0$ become occupied equally.
At the third resonance the latter state also becomes accessible
via $M= 0 \leftrightarrow M'= \pm 1/2$.
In the presence of QTM, $B_2 \neq 0$, two additional
resonances appear, see Fig. \ref{fig:smallS}(b), one with positive and
one with negative $dI/dV_{\text{bias}}$.
The appearance of negative $dI/dV_{\text{bias}}$ is related to a slow, 
spin-forbidden transition as follows.
For $B_2 \ll D$ the spin-projection $M$ is only approximately a
good quantum number, i.e. each eigenstate is a linear combination of
states $\lbrace | N , S, M + 2k \rangle \rbrace_{k=0,\pm 1, \pm 2,...}$, with
one coefficient ($k=0$) close to 1. 
In the $N=1$ excited state in addition to the state $M'=\pm 1/2$,
there is thus a small admixture $\propto B_2/D^{(1)}$
of state $M=\mp 3/2$.
The forbidden transition to the $N=0$ ground states composed mostly of $M
= \pm 2$ [Fig. \ref{fig:model}(a)] is now weakly allowed.
When it becomes energetically allowed the transition occurs with rate
$W_f \sim (B_2/D^{(1)})^2 \Gamma$ and the current is suppressed.
This is simply because the occupation of the $N=1$ excited states
reduces the occupations of the states which contribute most to the
transport current through (fast) spin-allowed transitions.
In contrast, the positive $dI/dV_{\text{bias}}$ line which appears
is \emph{not} related to any addition energy of the molecule.
The current increase  occurs when the state causing the negative differential 
conductance (NDC) above is depleted at higher bias via a spin-allowed transition
$M=\pm 1 \leftarrow M'=\pm \half$  (dashed in Fig. \ref{fig:model}).
The rate for this process is $W_a \sim \Gamma f(\Delta E-V_{bias}/2)$
where $\Delta E$ denotes the corresponding transition energy.
This depletion sets in when in- and out-going rates become equal i.e.
$W_a \sim W_f$.
Due to the small factor in $W_f$ this occurs already for $V_{bias}/2 <
\Delta E$ where $W_a \approx \Gamma \exp(-(\Delta E - V_{bias}/2)/T)$.
Equating the rates we obtain the resonance condition
$V_{bias}/2-\Delta E \propto  T \ln (D^{(1)}/B_2)$
which is substantially shifted from the position
expected naively ($V_{bias}/2=\Delta E$). The shift is linear in
temperature and logarithmical in the QTM amplitude~\cite{Bonet02note,Gol04}.
The shift with temperature can be larger than the thermal smearing as
illustrated in Fig. \ref{fig:smallS}(c).
Thus due to the asymmetry between electron tunneling rate constants
 intrinsic to an SMM, transport resonances appear even when the molecular
level is far away from the electrochemical potential.
The strong Coulomb charging effect and energy quantization on the molecule
are essential to this effect since they restrict the transport to
sequential tunneling through two charge states.
\\
For SMMs with large spin, $S^{(0)}=S^{(1)}+\half > 2$, the above
mechanism leads to \emph{oscillations} in the transport quantities, 
shown in Fig.~\ref{fig:bigS}(a)-(c). 
At low bias the states of the ``flatter'' $N=1$ parabola
become occupied via spin-forbidden transitions, see Fig. \ref{fig:bigS} (a)
and the current decreases.
The depletion of these states by spin-allowed transitions increases the
current again, Fig. \ref{fig:bigS} (b).
Due to the peculiar inverted parabolic energy dependence of the magnetic
excitations, 
this sequence is repeated whenever a new  $N=0$ excitation state
can be occupied, see Fig.~\ref{fig:model}(b)-(c).
With increasing bias the $N=0$ excitations are successively occupied
whereas the occupations of the $N=1$ excitations, and therefore also the
current \emph{oscillate}, see Fig. \ref{fig:bigS}(d).
Interestingly, all NDC resonances in  Fig.~\ref{fig:bigS}(a) and (b)
correspond to addition energies of the SMM (as in our previous example).
Most other resonances with positive differential conductance are fake
since they shift with $T$ and $B_2$. The \emph{shot-noise} also 
shows oscillations as function of $V_{\text{bias}}$:
in Fig. \ref{fig:bigS} (c) the Fano-factor $F=S/(2I)$ is plotted.
The periodic reductions of $F$ occur due to concerted
\emph{reductions of the noise} $S$ and simultaneous
\emph{enhancements of the current} $I$.
Their positions shift with  $T$ and $B_2$ and are fake.
The noise is super-poissonian, $F \gg 1$, due to the presence of 
slow and fast tunnel processes that give rise to large current
fluctuations~\cite{Belzig05prb}. 
Fig.~\ref{fig:bigS} (d) clearly shows that equal occupation of all
$N=1$ excitations associated with small spin-forbidden tunnel
constants reduces the current and simultaneously leads to 
stronger fluctuations. 
Depopulation of these states by a spin-allowed transition 
enhances the current and reduces the noise. 
Importantly, in Fig.~\ref{fig:bigS}(c)  the (white) lines of enhanced noise 
\emph{persist} in the Coulomb blockade regime, in contrast to the fake
(black) lines of noise reduction. 
The reason for this effect is that the noise in the Coulomb
blockade regime will only increase when more excitations lie in the
transport window as shown in~\cite{Belzig05prb}.   
Hence measuring shot-noise allows for an identification of fake resonance lines 
without changing the temperature 
(which may lead to unwanted changes in the molecular junction). 
We note that the above is valid also for a weaker magnetic distortion than 
$D^{(1)}/D^{(0)} < 1/2$ (which is the requirement for the
maximum number of oscillations to occur). \\
\emph{High-symmetry anisotropy. } When the dominant 
transverse anisotropy has a high symmetry, i.e. $B^{}_4 \gg B^{}_2/S^2$, 
and the \emph{ratio} $B^{}_4/D^{(N)}$ differs for $N=0$ and $N=1$ 
\emph{complete current blockade} may occur. 
This is shown in Fig.~\ref{fig:bigS_block} for $S^{(0)}=10$ and $S^{(1)}=9\half$.
For $B^{}_4/D^{(1)}\sim 1/S^2$ a level crossing occurs 
between ground and excited state in the charge sector $N=1$ [Fig. \ref{fig:bigS_level}]: 
the ground states change from a linear combination of
$\lbrace | \mp S^{(1)} \pm 4k \rangle_{z} \rbrace_{k=0,1,...}$
to a superposition of 
$\lbrace | \mp (S^{(1)} - 1) \pm 4k \rangle_{z} \rbrace_{k=0,1,..}$.
The latter states have very small tunneling overlap with the $N=0$
ground state which is a superposition of
$\lbrace | \mp S^{(0)} \pm 4k \rangle_{z} \rbrace_{k=0,1,...}$
for sufficiently small $B^{}_4/D^{(0)} \ll 1/S^2$.
Thus the transport suppression at low bias signals a high-symmetry QTM
perturbation.
It can also occur for constant $D^{(0)} \approx D^{(1)} \approx D$ when
$B_4$ changes from smaller than $D/S^2$ in one charge state to larger
than this value.
However, if the \emph{symmetry} of the QTM is also changed by the
charging, i.e. the low symmetry QTM 
 becomes important ($B_2 \sim B_4 S^2$) in one charge
state, the current blockade can be lifted, since the overlap of ground
states is restored.
This symmetry lowering may be expected when extra or deficit electrons
on the SMM are strongly localized on a particular metal ion
contributing to the total spin.
\\
\emph{Conclusion.} 
Transport spectroscopy of magnetic molecules is a
challenging task since typically many resonances are hidden by
spin-selection rules or do not correspond to addition energies of the molecule 
and shift with temperature. 
Measuring shot-noise allows for an identification of 
misleading excitations \emph{without changing the temperature}.
If the total spin values are known, a reconstruction
of the spectrum from the NDC excitations is possible. 
Then the fake resonances allow the determination of
the quantum tunneling parameter to logarithmic accuracy
even though it cannot be resolved directly from the thermally
broadened excitations.
Finally, we showed that the transport is even sensitive to the
symmetry of the magnetic anisotropy of the SMMs.
The link we established between transport effects and spin-Hamiltonian
parameters may be extended down to the microscopic details of magnetic
molecules with further input from ab-initio calculations and energy
spectroscopy on \emph{charged} states of SMMs.\\
We acknowledge discussions with H. Heersche, J. Kortus, 
H. van der Zant, and R. Sessoli, and financial support
through the Virtual Institute ``Functional Molecular Systems for
Information Technology'' and
the EU RTN Spintronics program HPRN-CT-2002-00302.
\bibliographystyle{apsrev}
\bibliography{paper}

\end{document}